\documentclass[a4paper,11pt]{article}
\pdfoutput=1 
\usepackage{jheppub} 
\usepackage[T1]{fontenc} 
\usepackage{color}

\def \be {\begin{equation}}
\def \ee {\end{equation}}
\def \ba {\begin{aligned}}
\def \ea {\end{aligned}}
\def \bea {\begin{eqnarray}}
\def \eea {\end{eqnarray}}

\title{The radial distribution function reveals the underlying mesostructure of the AdS black hole}

\author[a,b]{Conghua Liu,}
\author[c,*]{Jin Wang \note[*]{Corresponding author}}
\affiliation[a]{College of Physics, Jilin University, Changchun 130022, China}
\affiliation[b]{State Key Laboratory of Electroanalytical Chemistry, Changchun Institute of Applied Chemistry, Chinese Academy of Sciences, Changchun 130022, China}
\affiliation[c]{Department of Chemistry and Department of Physics and Astronomy, State University of New York at Stony Brook, Stony Brook, New York 11794, USA}

\emailAdd{liuch20@mails.jlu.edu.cn}
\emailAdd{jin.wang.1@stonybrook.edu}

\abstract{Based on the equations of state, one can infer the underlying interaction potentials among the black hole molecules in the case of Schwarzschild-AdS and charged AdS black holes. The microscopic molecules with the interaction potential arrange in a specific way to form the mesostructure, whose size is between the macro (black hole system) and the micro (black hole molecules). As a result, the mesostructure leads to the emergence of the macroscopic phase. However, the information about the mesostructure of the AdS black hole are still elusive. In this paper, the radial distribution function is introduced to probe the mesostructure of the AdS black hole. We find that the mesostructure of the Schwarzschild-AdS black hole behaves as the ideal gas when the temperature is high. Furthermore, we find the mesostructure for the liquid-like (gas-like) phase of the small (large) charged AdS black hole. A sudden change of the mesostructure emerges from the liquid-like phase to the gas-like phase when the charged AdS black hole undergoes a phase transition from the small to large black hole, consistent with the viewpoint that the phase transition of the charged AdS black hole is reminiscent of that of the vdW fluid. This study provides a new angle towards understanding the black hole from its mesostructure.}

\begin{document}

\maketitle

\flushbottom

\section{Introduction}

The black hole is a mysterious and fascinating object. The classical black hole is a prefect absorber but emits nothing. However, since the quantum effects were considered, leading to the famous Hawking radiation, the black hole has been found to possess temperature with a black body spectrum~\cite{AA,AB,AC}. Namely, the black hole is not only a gravitational dynamical system, but also a thermodynamic system~\cite{AD}. Thereafter, the thermodynamics has been generally used to study the properties of the black hole~\cite{AE,AF,AH,AI,AJ,AK,AL,AM,AN,AO,AP,AQ,AR,AT}. One interesting field is the phase transition of the black holes. The Hawking-page transition is a first order phase transition between the radiation and the large Schwarzschild anti-de Sitter black hole, which can be interpreted as the confinement/deconfinement phase transition of gauge field in the context of AdS/CFT correspondence~\cite{AE,AF}. After treating the cosmological constant as the thermodynamic pressure, the analogies between the charged AdS black hole and the vdW fluid have been established~\cite{AJ,AK,AL,AM}. Furthermore, the dynamical processes of the phase transition in Schwarzschild-AdS black hole and charged AdS black hole have also been studied~\cite{AP,AQ,AR}.

However, there are still some differences between the ordinary thermodynamic system and the black holes with a strong gravity. One important difference is the entropy of the black hole is proportional to the area rather than the volume. Such a unique property has attracted much attention, even though each theory has its own limit~\cite{AU,AV,AW,AX}. 

From the view of the Boltzmann, “If you can heat it, it has microscopic structure". The black hole, as a thermodynamic system, should also have its microscopic structure. In Ref.~\cite{AY}, Wei and Liu proposed that the AdS black hole had microstructure named black hole molecules and the number density was introduced to measure the microscopic degrees of freedom. Different from the ordinary thermodynamic system where the macroscopic thermodynamic quantities can be derived from the microscopic structure by the statistical physics, for the black hole, we can only conjecture the microscopic structure from the known information about the macroscopic quantities. In Ruppeiner geometry, the sign of scalar curvature provides us information about the interaction type of the black hole molecules~\cite{AZ,BA,BB,BD,BE,BF,BG,BH,BP,BQ}. Namely, the positive or negative scalar curvature corresponds to the repulsive and the attractive interaction respectively. Very recently, the interaction potentials of the AdS black hole molecules have been investigated, which shows the Lennard-Jones potential as a suitable candidate to describe the interaction among the black hole molecules~\cite{BF,BG,BH}. All these studies provide better understanding towards the microscopic structure of the AdS black hole. However, the studies on the arrangement of the black hole molecules, i.e. the mesostructure of the AdS black hole, are still absent. In this paper, we try to probe the mesoscopic structure of the AdS black hole.


\section{Schwarzschild-AdS black hole}
\subsection{The interaction potential of the Schwarzschild-AdS black hole}
The metric of four-dimensional Schwarzschild-AdS (SAdS) black hole is written as:
\be
\ba
\label{eq:1}
ds^2=-f(r)dt^2+\frac{dr^2}{f(r)}+r^2d{\Omega}^2,
\ea
\ee
where $f(r)$ is given by
\be
\ba
\label{eq:2}
f(r)=1-\frac{2M}{r}+\frac{r^2}{L^2}.
\ea
\ee
Here, $M$ is the mass and $L$ is the AdS curvature radius which is associated with the negative cosmological constant $\Lambda$ by $L=\sqrt{\frac{-3}{\Lambda}}$.

The Hawking temperature is given as
\be
\ba
\label{eq:3}
T=\frac{f'(r_+)}{4\pi}=\frac{1}{4{\pi}r_{+}}(1+\frac{3r_{+}^2}{L^2}),
\ea
\ee
where $r_+$ is the horizon radius determined by $f(r_+)=0$.

After considering the cosmological constant as the analogue of the thermodynamic pressure by~\cite{AO,AJ,AK,AL,AM}:
\be
\ba
\label{eq:4}
P=-\frac{\Lambda}{8\pi},
\ea
\ee
the temperature equation~(\ref{eq:3}) can be rewritten as~\cite{AM}:
\be
\ba
\label{eq:5}
P=\frac{T}{v}-\frac{1}{2\pi v^2},
\ea
\ee
where $v$ is the specific volume of the black hole molecules with the definition $v=2r_+$~\cite{AL,AM}. 

Actually, Eq.~(\ref{eq:5}) is the equation of state for the SAdS black hole, which is similar to the equation of state for the vdW fluid with the form:
\be
\ba
\label{eq:6}
P=\frac{T}{v-b}-\frac{a}{v^2},
\ea
\ee
where $b$ is the nonzero size of the fluid molecules and $a$ is the interaction between these molecules. 

\begin{figure}[b]
\centering
\includegraphics[width=0.46\textwidth]{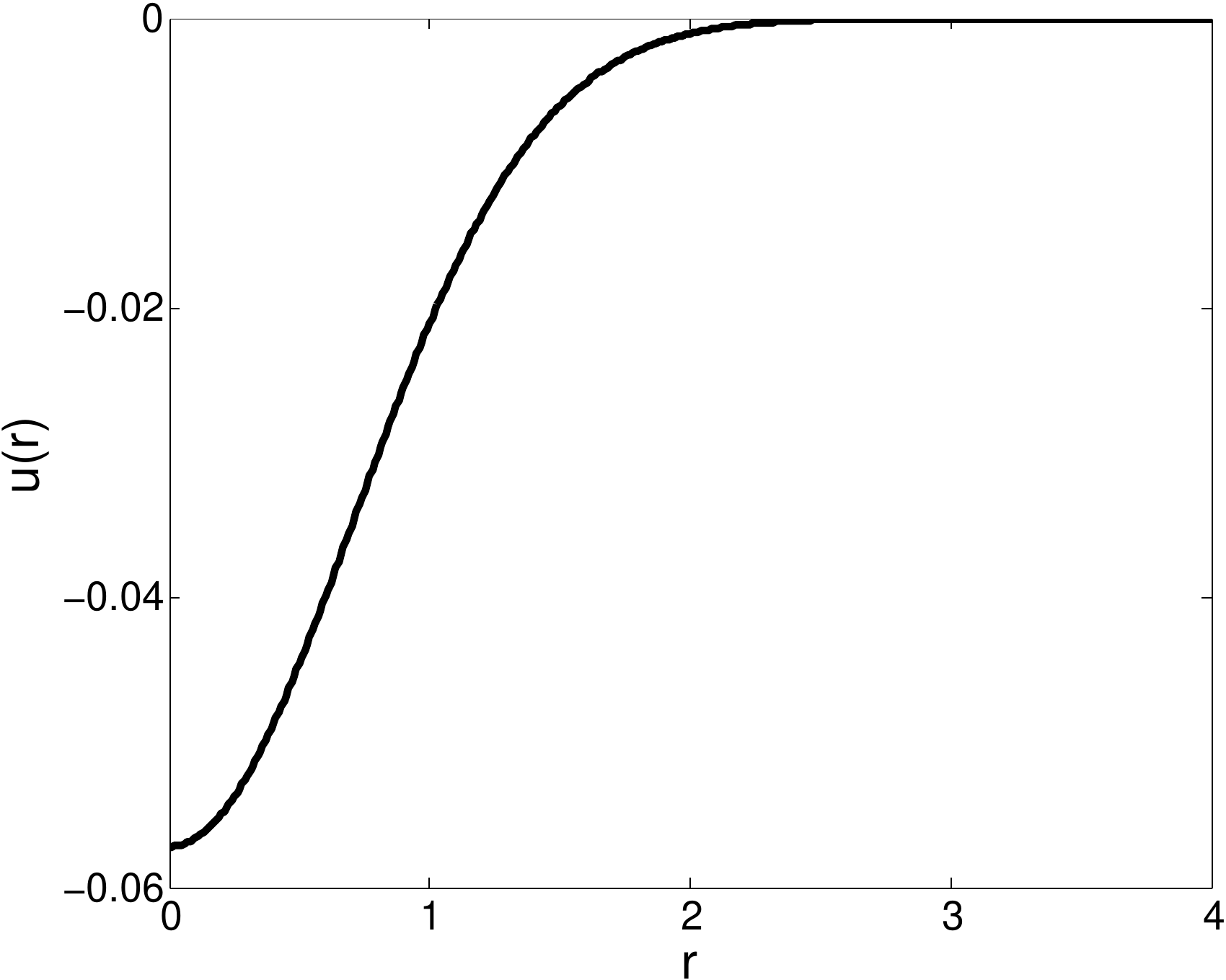}
\caption{The interaction potential of the Schwarzschild-AdS black hole.}
\label{Fig1}
\end{figure}

The Eq.~(\ref{eq:5}) can be associated to Eq.~(\ref{eq:6}) by taking 
\be
b=0, \qquad a=\frac{1}{2\pi}.
\ee

It indicates the volume of the SAdS black hole molecules is zero. The Ruppeiner geometry provides information about the characteristics of the interaction among the black hole molecules~\cite{AZ,BA,BB,BD,BE,BF,BG,BH,BP,BQ}. Namely, when the scalar curvature $R>0$ ($R<0$), the repulsive (attractive) interaction dominates. While for $R=0$, the interaction vanishes. In Ref.~\cite{BP,BQ}, the scalar curvature of SAdS black hole is found to be negative, which means the attractive interaction dominates for the SAdS black hole. It is worth noting that the form of the interaction potential is not unique. However, if we choose a special form of the potential which is satisfied with the Ruppeiner geometry, the potential can be determined uniquely. Different from the previous Lennard-Jones potential in~\cite{BH}, we consider a new form of the effective interaction potential as:
\be
\ba
\label{eq:7}
u(r)=-u_0\exp{[-(\frac{r}{r_0})^2]},
\ea
\ee
where $u_0$ reflects the interaction strength of the black hole molecules, and $r_0=1$ is introduced for dimensional purpose. Under the mean field approximation, $u_0$ can be calculated by the equation~\cite{BI}:
\be
\ba
\label{eq:8}
a=-2\pi\int_0^{\infty}u(r)r^2 dr,
\ea
\ee
which yields $u_0=\frac{1}{{\pi}^{\frac{5}{2}}}$. The interaction potential is displayed in Fig.~\ref{Fig1}, from which we can clearly see that the attractive interaction dominates.


\subsection{The radial distribution function of the Schwarzschild-AdS black hole}

After the interaction potential among the black hole molecules is obtained, we can investigate the mesostructure of the SAdS black hole by introducing the radial distribution function $g(r)$, which is defined as~\cite{BJ,BL}:
\be
n^{(2)}=n^2 g(r),
\ee
where $n$ is the number density of the black hole molecules which measures the microscopic degrees of freedom of the black hole with the definition $n=\frac{1}{v}=\frac{1}{2r_+}$\cite{AY}, $n^{(2)}(r)$ is the probability density that two molecules will be found at distance $r$. The radial distribution function $g(r)$ is proportional to the probability density that a molecule may be found at a distance $r$ from the reference molecule. If the reference molecule is located at the origin with probability density $n$, then the probability density of finding a molecule at distance $r$ from the origin is given by $ng(r)$. $g(r)$ is a measure of the spatial distribution for the interaction among the molecules. For independent molecules, $g(r)=1$ and $n^{(2)}(r)=n^2$. But for the correlated molecules, $g(r)$ is not usually a constant. 

\begin{figure}[b]
\centering
\includegraphics[width=0.46\textwidth]{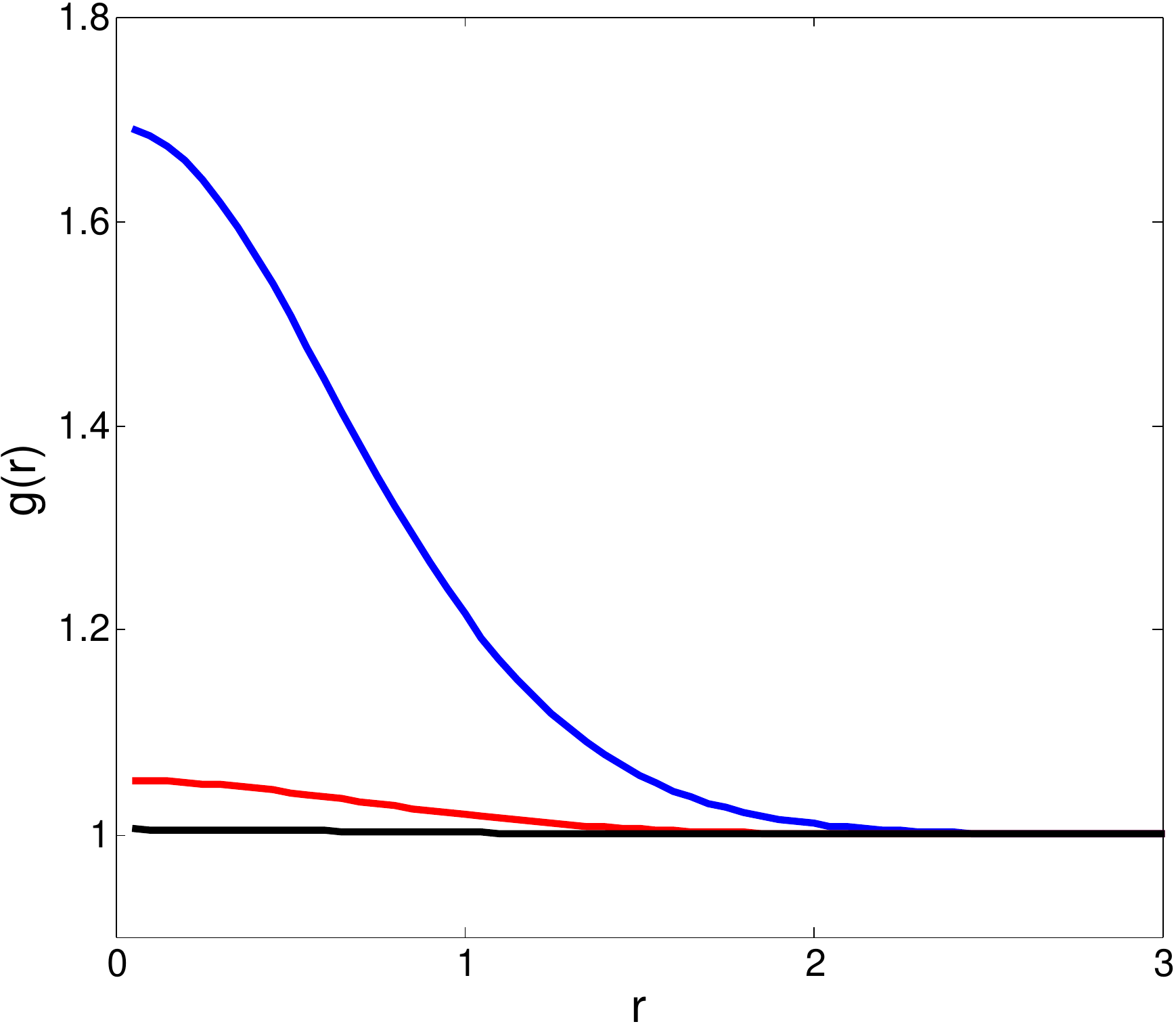}
\caption{The radial distribution functions $g(r)$ of the Schwarzschild-AdS black holes at $P=0.001$ and various temperatures. The values of the temperatures from top to bottom are set as $T=0.1$, $T=1$ and $T=10$. The large Schwarzschild-AdS black holes are globally stable at these temperatures and pressure.}
\label{Fig2}
\end{figure}

After we choose a reference molecule, the number of molecules in a sphere of radius $R$ with the reference molecule as the center can be given by~\cite{BJ,BL}:
\be
\ba
\label{eq:99}
N(r)=n\int_0^R 4\pi r^2 g(r) dr.
\ea
\ee
This indicates that the information about the molecular arrangement in radial distribution can be derived from the radial distribution function directly.

The exact calculation of $g(r)$ is a difficult task, fortunately, the approximation methods have been developed and used to solve many problems successfully~\cite{BJ,BL,BM,BN,BO,BS}. The Percus-Yevick approximation is one of the best approximation methods, in which Percus and Yevick used the collective coordinate techniques to simplify the calculations~\cite{BS}. By means of the PY approximation, the $g(r)$ is determined by~\cite{BM,BN,BS}:
\be
\ba
\label{eq:10}
h'(r)=1-2\pi n\int_{0}^{\infty}\{1-\exp[-\beta u(s)]\}h(s)\{h(s+r)\exp[-\beta u(s+r)]+\\
h(|s-r|)\exp[-\beta u(|s-r|)]\frac{s-r}{|s-r|}-2s\}ds,
\ea
\ee

\be
\ba
\label{eq:11}
h(r)=\int_{0}^{r}h'(t)dt,
\ea
\ee
\be
\ba
\label{eq:12}
h(r)=rg(r)\exp[\beta u(r)].
\ea
\ee

\begin{figure}[b]
\centering
\includegraphics[width=0.96\textwidth]{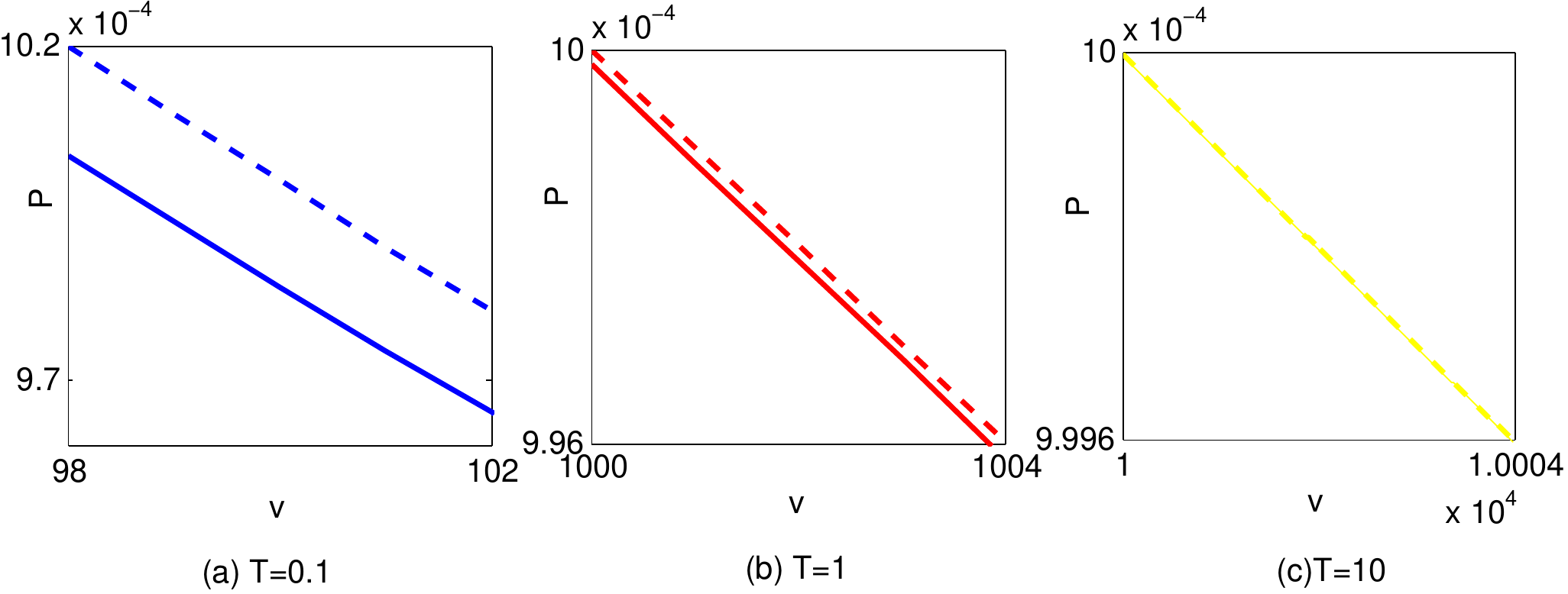}
\caption{The equations of state of the Schwarzschild-AdS black hole and the ideal gas at the same temperatures taken in Fig.~\ref{Fig2}. In each subfigure, the solid curve represents the equation of state of the Schwarzschild-AdS black hole, and the dashed curve represents that of the ideal gas. }
\label{Fig3}
\end{figure}

Here, $n$ is the number density and $\beta=\frac{1}{kT}$, where $k$ is the Boltzmann constant taken as $1$. $h(r)$ can be calculated numerically by iteratively solving the equation (\ref{eq:10}) and (\ref{eq:11}), and therefore $g(r)$ can be obtained by the equation (\ref{eq:12}) (see Appendix.~\ref{sec:Det}). In Fig.~\ref{Fig2}, we have displayed the radial distribution functions of the SAdS black holes at different temperatures. Different from the interaction potential, which reflects the local information about how two molecules influence each other, the radial distribution function reflects the global information about how these molecules arrange, thereby how the black hole state behaves can be inferred.  We observe that the the radial distribution function of the SAdS black hole gradually becomes a constant $1$ as the temperature increases. This indicates that the interaction will disappear and the molecules of the SAdS black hole become independent with each other when the temperature is high. Namely, the mesostructure of the SAdS black hole in high temperature behaves as the ideal gas. Such behaviors are consistent with Fig.\ref{Fig3}, in which we have plotted the equations of state of both the Schwarzschild-AdS black hole and the ideal gas at the same temperatures taken in Fig.~\ref{Fig2}. We can clearly observe the curves of SAdS black hole and ideal gas will gradually overlap when the temperature increases, which means the equation of state of the SAdS black hole behaves as that of the ideal gas when the temperature increases. The results are consistent with Fig.~\ref{Fig2}, in which the radial distribution function of SAdS black hole is shown to  behave like the ideal gas when the temperature is high.


\section{The charged AdS black hole}
\subsection{The interaction potential of the charged AdS black hole}
When we take account of the charge $Q$ into the Schwarzschild-AdS black hole, we can obtain the metric of the charged AdS black hole as Eq.~(\ref{eq:1}), for which $f(r)$ is given by
\be
\ba
\label{eq:13}
f(r)=1-\frac{2M}{r}+\frac{Q^2}{r^2}+\frac{r^2}{L^2}.
\ea
\ee

The equation of state of the charged AdS black hole is written as~\cite{AL,AM}:
\be
P=\frac{T}{v}-\frac{1}{2\pi v^2}+\frac{2Q^2}{\pi v^4}.
\ee

There are many similarities between the charged AdS black hole and the vdW fluid. In the vdW fluid, the equation of state Eq.(\ref{eq:6}) shows both liquid and gas phases with the possibility of a first order phase transition in between. But at the critical point $(P_c, T_c, v_c)=(\frac{a}{27b^2}, \frac{8a}{27b}, 3b)$, which is determined by $(\frac{\partial P}{\partial v})_T=0$ and $(\frac{\partial^2 P}{\partial v^2})_T=0$, the phase transition becomes second order. Analogously, the charged AdS black hole also has two stable phases, i.e. the small and large black holes, which are characterized by the radii of the event horizons. Between the small and large black holes, a first order phase transition can occur, and changes to the second order at the critical point $(P_c, T_c, v_c)=(\frac{1}{96\pi Q^2}, \frac{\sqrt{6}}{18\pi Q}, 2\sqrt{6}Q)$~\cite{AL,AM}. Comparing the critical point of the charged AdS black hole with that of the vdW fluid, one can obtain~\cite{AL}:
\be
a=\frac{3}{4\pi}, \qquad  b=\frac{2\sqrt{6}}{3}Q.
\ee

\begin{figure}[htp]
\centering
\includegraphics[width=0.46\textwidth]{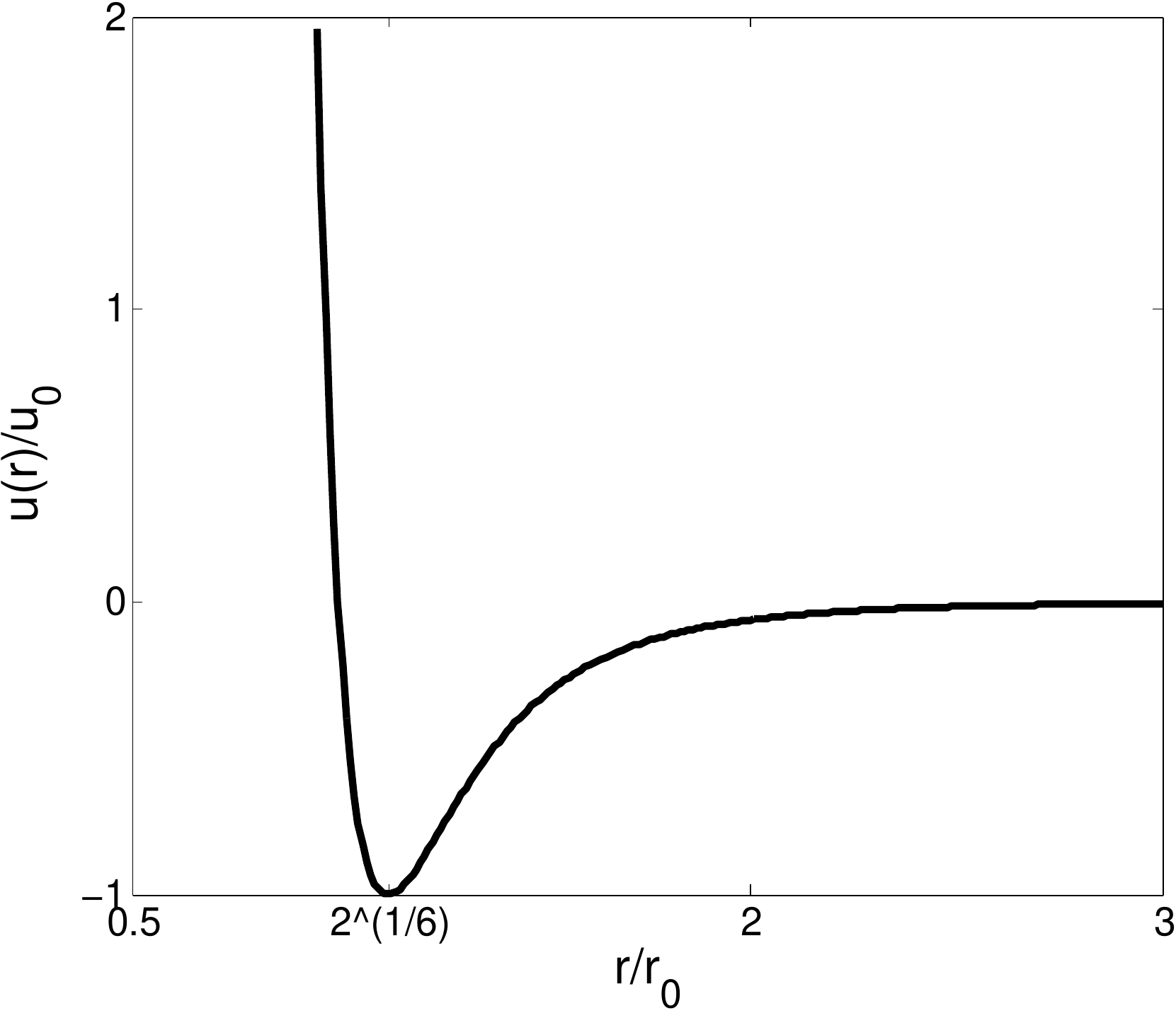}
\caption{The Lennard-Jones 12-6 potential.}
\label{Fig4}
\end{figure}

In the van der Waals theory, the Lennard-Jones potential provides a good description to the interaction of the vdW fluid molecules~\cite{BI}. In Ref.~\cite{BD}, the scalar curvature of the charged AdS black hole is found to be positive or negative in small or large volume, which corresponds to a repulsive and attractive interaction dominates respectively. Such characteristic can be well described by the LJ potential with the form~\cite{BF,BG,BH}:
\be
u(r)=4u_0[(\frac{r_0}{r})^{12}-(\frac{r_0}{r})^6],
\ee
where $r=r_0$ corresponds to $u(r)=0$, and $u(r)$ takes the minimum value when $r=r_1=2^{\frac{1}{6}}r_0$. In Fig.~\ref{Fig4}, we have plotted the LJ potential.

From Ref.~\cite{BD}, we know that the scalar curvature vanishes when $v=2\sqrt{2}Q$. We note that such a specific volume is not completely occupied by a sphere, but should be composed of the maximum volume sphere and the interval. We consider the closest arrangement of these maximum volume spheres, and find that the volume of the sphere occupies $\frac{\sqrt{2}\pi}{6}$ of the specific volume (see the Appendix.~\ref{sec:Cal}). From Fig.~\ref{Fig4}, we know that two molecules will not influence each other if the distance of the two molecules is $r_1$, i.e. the minimum point of the LJ potential, which should coincide with the point of vanishing scalar curvature. Thus, we can obtain:
\be
\ba
\label{eq:33}
\frac{\sqrt{2}\pi}{6}*2\sqrt{2}Q=\frac{4}{3}\pi (\frac{r_1}{2})^3.
\ea
\ee

Without loss of the generality, we take $Q=1$ for the calculations in the whole paper, and Eq.~(\ref{eq:33}) yields $r_0=\frac{r_1}{2^{\frac{1}{6}}}=1.414$.

We assume the distance between two molecules can not be smaller than the diameter of the molecule, also called “the hard sphere approximation". As said in ~\cite{BH,BI}, $b$ is the minimum value of the specific volume. When $v=b$, it means the space is completely occupied by the molecules~\cite{BH,BI}. For the minimum value of the specific volume $b$, there is also $\frac{\sqrt{2}\pi}{6}$ of it occupied by the sphere, namely, the volume of the molecule. The diameter of the molecule is denoted as $d$, and we have:
\be
\frac{\sqrt{2}\pi}{6}b=\frac{4}{3}\pi (\frac{d}{2})^3,
\ee
yielding $d=1.322$.

Then, the parameter $u_0$ can be calculated by the mean field approximation~\cite{BI}:
\be
a=-8\pi u_0 \int_d^{\infty}[(\frac{r_0}{r})^{12}-(\frac{r_0}{r})^6]r^2 dr.
\ee
This equation yields $u_0=0.01645$.


\subsection{The radial distribution function and the phase transition of the charged AdS black hole}

The coexistence curve of the charged AdS black hole satisfies the equation~\cite{AN}:
\be
{\tilde{T}}^2=\frac{\tilde{P}(3-\sqrt{\tilde{P}})}{2},
\ee
where $\tilde{T}=\frac{T}{T_c}$ and $\tilde{P}=\frac{P}{P_c}$. Below the critical point $(P_c, T_c, v_c)=(\frac{1}{96\pi Q^2}, \frac{\sqrt{6}}{18\pi Q}, 2\sqrt{6}Q)$, it is found that the number density will have a discontinuous change when the small black hole (SBH) crosses the coexistence curve to become the large black hole (LBH)~\cite{AY}. Indeed, when we choose $T=0.5T_c$ and $P=0.1955P_c$ on the coexistence curve, the number density suffers a change from $n=0.418$ (SBH) to $n=0.044$ (LBH).

\begin{figure}[t]
\centering
\includegraphics[width=0.48\textwidth]{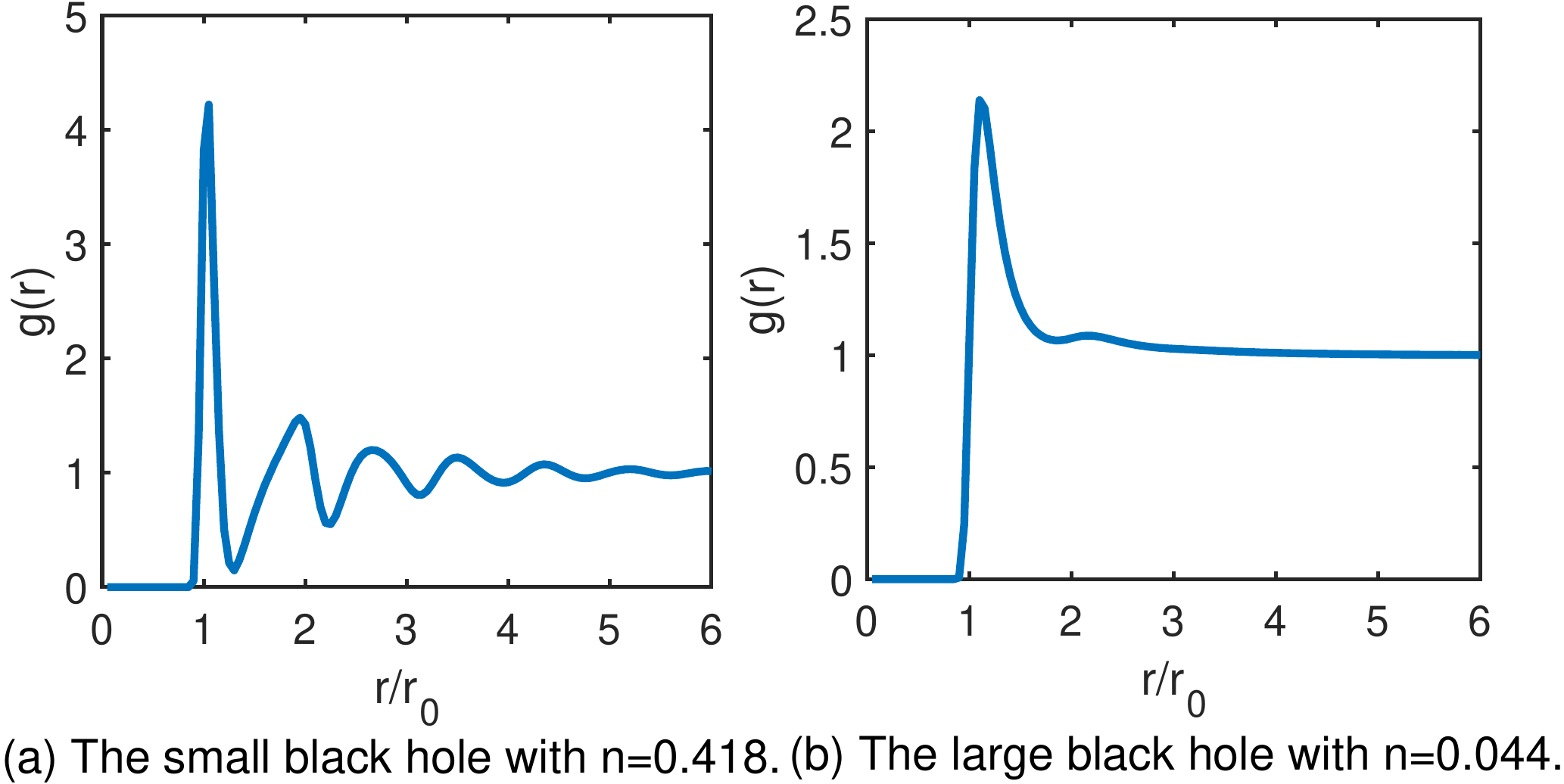}
\caption{The radial distribution functions $g(r)$ of the small and large black holes at $T=0.5T_c$ and $P=0.1955P_c$. When the small black hole crosses the coexistence curve to become the large black hole, there is a sudden change of the mesoscopic structure from subfigure $(a)$ (left) to subfigure $(b)$ (right).}
\label{Fig5}
\end{figure}

\begin{figure}[t]
\centering
\includegraphics[width=0.48\textwidth]{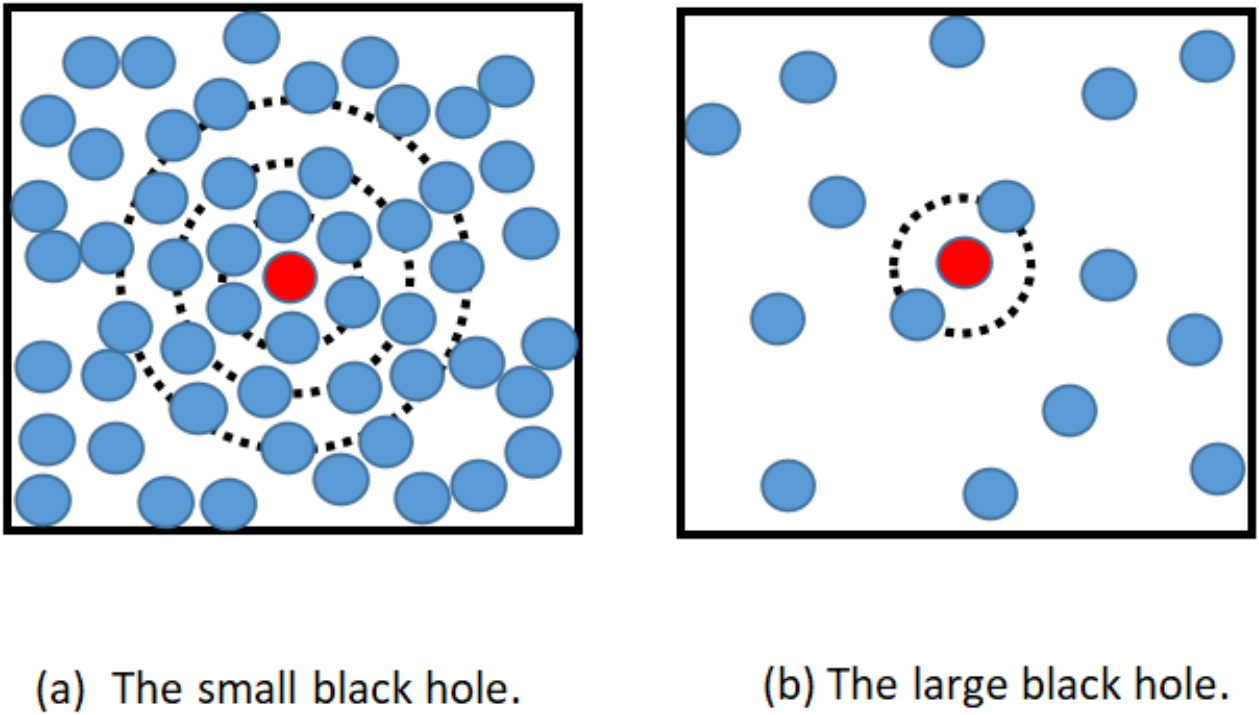}
\caption{Sketches of the mesoscopic structures for the charged AdS black holes. (a) The mesoscopic structure of the small black hole behaves as the liquid. (b) The mesoscopic structure of the large black hole behaves as the gas. }
\label{Fig6}
\end{figure}

When such a phase transition undergoes, we show the change of the radial distribution function in Fig.~\ref{Fig5}. The large $r$ corresponds to $g(r)=1$, which means if the distance between two molecules is long, the interaction will disappear and the molecules become independent with each other. As $r$ is smaller than the diameter of the black hole molecule, $g(r)=0$ and there is no possibility to find a molecule in such a distance. The maximal peak of the radial distribution approximately corresponds to the minimum of the potential in Fig.~\ref{Fig4}, which means the black hole molecules will be located  with the largest probability at the position where the attractive force and the repulsive force are balanced. For the small black hole, the radial distribution function has a certain number of damped peaks, and the positions of these peaks provide information about the separation of the molecules in the mesostructure directly. The mesoscopic structure of the small black hole shows a short-range order, which behaves as the liquid (We recommend Figure $6.5$ in~\cite{BJ} to serve as a dictionary, so that one can clearly see visually how the $g(r)$ behaves after a comparison). However, for the large black hole, these damped peaks disappear and the radial distribution function rapidly becomes constant $1$ as the distance increases, which is similar to the mesoscopic structure of the gas. When the small black hole goes across the coexistence curve to become the large black hole, there is not only a discontinuous change of the number density, but also a sudden change of the mesostructure from the liquid-like type to the gas-like type, which is in agreement with the viewpoint that the phase transition of the charged AdS black hole is analogous to that of the vdW fluid.

In Fig.~\ref{Fig6}, we have displayed the sketches of mesostructures for the small and large black holes. The red spheres represent the reference molecules, and the dotted circles from inside to outside represent the locations of the first, second and third peaks of the radial distribution function, where the molecules lie at the first peak with the largest possibility. Since $g(r)$ of the small black hole shows a short-range order, similar to the liquid, the molecules in subfigure $(a)$ arrange orderly near the reference molecule, and turn to be unordered as the distance increases. For the large black hole, without a short-range order, the molecules in subfigure $(b)$ arrange in a disordered fashion, which behave as the gas. It is worth noting that the exact diagram of the radial arrangement in three-dimensional space for the black hole molecules can be drawn based on Eq.~(\ref{eq:99}).

Furthermore, the introduction of the radial distribution function can associate the macroscopic thermodynamic quantities with the mesostructure, and some useful thermodynamic quantities can be calculated from $g(r)$ by a simple formulae, such as~\cite{BM,BN}:
\be
\frac{\beta E}{N}=\frac{3}{2}+2\pi\beta n\int_{0}^{\infty}u(r)g(r)r^{2} dr,
\ee
\be
\frac{\beta P}{n}=1-\frac{2\pi\beta n}{3}\int_{0}^{\infty}u'(r)g(r)r^{3} dr.
\ee


\section{Conclusion}
In this paper, we have investigated the mesostructures of the Schwarzschild-AdS black hole and the charged AdS black hole. Based on the equation of state and the Ruppeiner geometry, we obtain the interaction potential among the black hole molecules. Aiming at understanding the mesostructure of the AdS black hole, we introduce the radial distribution function, which provides the information about molecular arrangement in the AdS black hole. We find that the mesoscopic structure of the Schwarzschild-AdS black hole behaves like the ideal gas when the temperature is high. For the charged AdS black hole, there is a sudden change from the liquid-like structure to the gas-like structure when the small black hole switches to the large black hole, which are consistent with the viewpoint that the phase transition of charged AdS black hole is of the vdW type. Furthermore, the radial distribution function also provides a bridge between the macroscopic thermodynamic quantities and the mesostructure, which gives us a simple formulae for getting the thermodynamic quantities. 

In condensed matter physics, the radial distribution functions of different materials can be obtained from experiments. The scattering experiments can measure the scattering structure function $S(Q)$, then the molecular pair distribution function (PDF) analysis, which is a Fourier analysis of the scattering data, can be used to get the radial distribution function. For the real black hole, it is currently impossible to measure the scattering structure function experimentally. However, it will be interesting to proceed such a scattering experiment in the laboratory simulation of the black hole, such as the acoustic black hole~\cite{BT,BU,BV,BW}. Finally, there are also many attractions to calculate the interaction potentials and the radial distribution functions for other interesting black holes to understand their mesostructures. These deserve the future studies.


\section*{Acknowledgments}
C. H. Liu thanks Wen-Ting Chu, Hong Wang, He Wang, Bo-Han Cao and Chao Yang for the helpful discussions. C. H. Liu thanks the support in part by the National Natural Science Foundation of China Grant No. 21721003, No. 32000888 and No. 32171244.

\vskip 1cm
\vfill
\appendix
\section{Details about the numerical iteration}
\label{sec:Det}

In order to obtain $g(r)$, we have used the PY approximation. The solution of the equation (\ref{eq:10}) and (\ref{eq:11}) is calculated by the iteration until $h(r)$ is self-consistent. In generally, the direct iteration of $h(r)$ will not give a convergent solution. We use a trick recommended in some studies~\cite{BM,BN}. If a first input value of $h(r)$ is denoted as “$h_{in}(old)$", and the resultant output value is denoted as “$h_{out}$", then the new input “$h_{in}(new)$" is given by,
\be
h_{in}(new)=\alpha h_{in}(old)+(1-\alpha)h_{out},
\ee
where $0<\alpha<1$. For a high density, a large value of $\alpha$ is required, and the speed of convergence will become slow. In the main text, we have used $\alpha=0.5$, $0.5$ and $0.99$, respectively for the calculations of Schwarzschild, large charged and small charged AdS black holes. The convergence criterion is given by:
\be
|h_{in}(new)-h_{in}(old)|< 10^{-5}.
\ee

Correspondingly, in order to satisfy the convergence criterion, the numbers of iteration are taken as $30$, $100$ and $6000$.

Before the calculations of $g(r)$ for the AdS black holes, we have calculated $g(r)$ of the Lennard-Jones potential with the condition in~\cite{BN} to examine the accuracy of our procedure, which gives a less than $1\%$ error compared with the data of TABLE IV. M in~\cite{BN}. Then, we use the procedure for the calculations of the AdS black holes.

\section{The calculations of $\frac{\sqrt{2}\pi}{6}$}
\label{sec:Cal}

\begin{figure}[htp]
\centering
\includegraphics[width=0.3\textwidth]{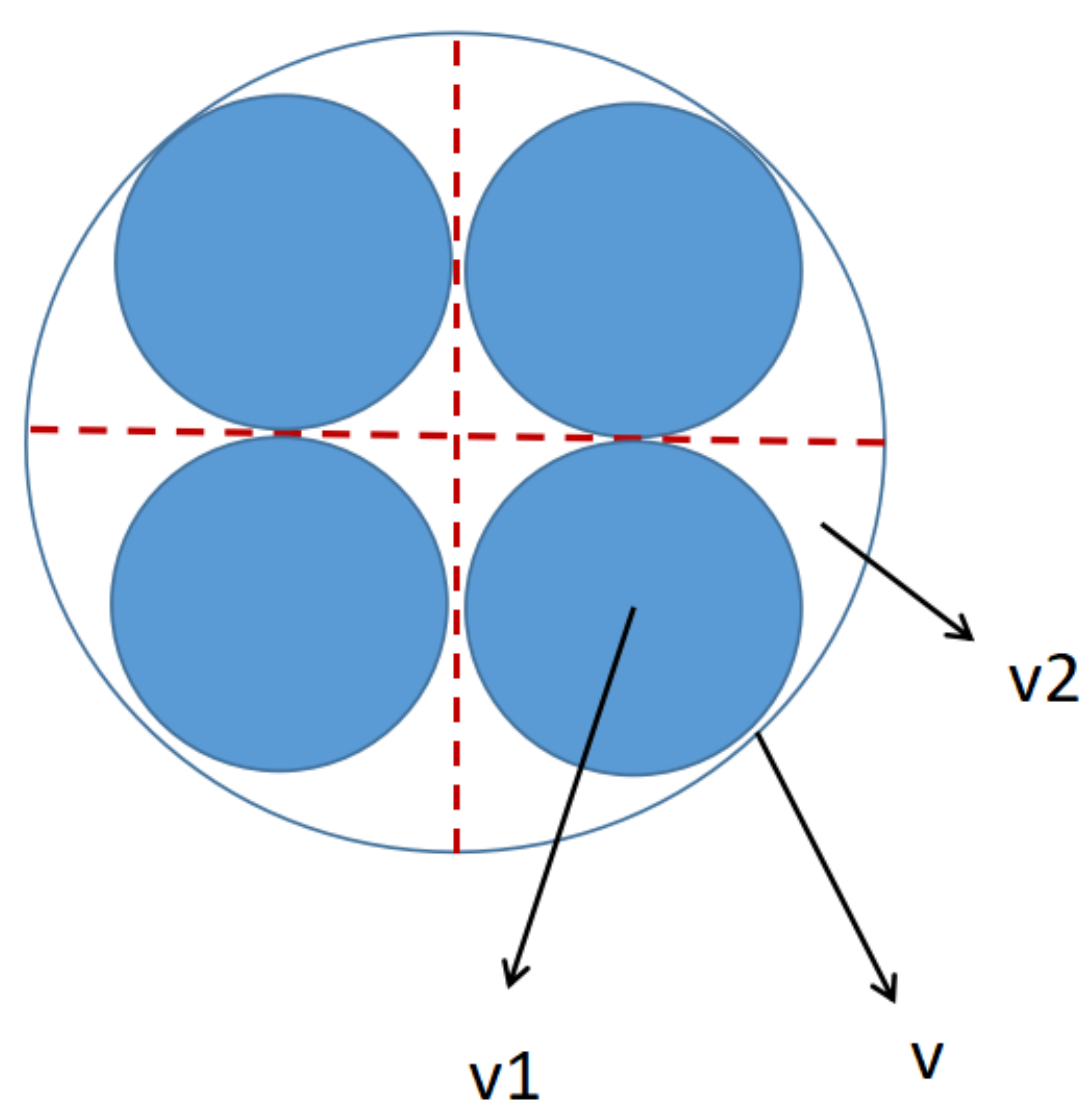}
\caption{The specific volume is composed of the maximum volume sphere and the interval.}
\label{Fig7}
\end{figure}

As shown in Fig.~\ref{Fig7},  a sphere can not be occupied by a series of small spheres with no interval. This means:

\be
V=Nv=N(v_1+v_2),
\ee
where $V$ is the volume of the black hole, $N$ is the number of the black hole molecules, and $v$ is the specific volume. $v$ is composed of two parts: One is the part of the maximum volume sphere denoted as $v_1$, the other is the part of the interval denoted as $v_2$.

These maximum volume spheres are closest to arrange. There are two kinds of the closest arrangement of the spheres, i.e. the face-centred cubic packing and the hexagonal closest packing, and they have the same rate occupied in the cubic by the sphere. We consider the face-centred cubic packing for the arrangement of the maximum volume spheres, for which each vertex angle and center of the face has a sphere. In Fig.~\ref{Fig8}, we have displayed the face-centred cubic packing, the red spheres and the yellow spheres respectively locate at the vertex angles and the centers of the faces of the cubic. We assume that the side length of the cubic is $2l$, and the volume of the cubic is $8l^3$. The diameter of maximum sphere is $\sqrt{2}l$, then the volume of the sphere is given by:
\begin{figure}[b]
\centering
\includegraphics[width=0.3\textwidth]{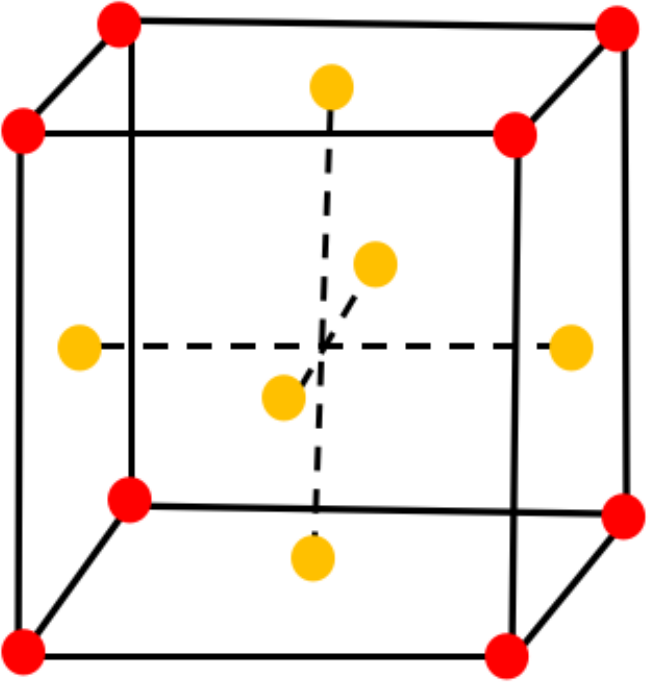}
\caption{The face-centred cubic packing.}
\label{Fig8}
\end{figure}
\be
v_1=\frac{4}{3}\pi (\frac{\sqrt{2}l}{2})^3=\frac{\sqrt{2}\pi l^3}{3}.
\ee

In such a cubic, the number of the sphere is:
\be
\frac{1}{8}{\times}8+\frac{1}{2}{\times}6=4.
\ee

Therefore, we can calculate the rate of the cubic occupied by the maximum volume sphere:
\be
\frac{4\times\frac{\sqrt{2}\pi l^3}{3}}{8l^3}=\frac{\sqrt{2}\pi}{6},
\ee
which is also the rate of the specific volume occupied by the sphere with maximum volume.

\end{document}